\documentclass[useAMS,usenatbib]{mn2e}
\usepackage{epsfig}
\usepackage{amsmath}
\usepackage{verbatim}
\usepackage{graphicx}

\title[Rotational fission of Trans-Neptunian Objects. The case of
Haumea.]{Rotational fission of Trans-Neptunian Objects. The case of Haumea.}
\author[J.L. Ortiz et al]
{J.L. Ortiz$^{1}$\thanks{E-mail:
ortiz@iaa.es},
A. Thirouin$^{1}$, A. Campo Bagatin$^{2,3}$, R. Duffard$^{1}$, J.
Licandro$^{4}$,
\newauthor
 D.C. Richardson$^{5}$, P. Santos-Sanz$^{1,6}$, N. Morales$^{1}$ and P.G.
Benavidez$^{2,3}$\\
$^{1}$Instituto de Astrof\'{\i}sica  de Andaluc\'{\i}a - CSIC, Apt
3004, 18008  Granada,  Spain\\
$^{2}$Departamento de Fisica, Ingenieria de Sistemas y teoria de la
Se\~{n}al, Universidad de Alicante, PO Box 99, 03080 Alicante, Spain\\
$^{3}$Instituto de F\'{\i}sica Aplicada a las Ciencias y la Tecnolog\'{\i}a,
Universidad de Alicante, PO Box 99, 03080 Alicante, Spain\\
$^{4}$Instituto de Astrofisica de Canarias, C/ Via Lactea sn La Laguna,
Tenerife, Spain\\
$^{5}$Department of Astronomy, University of Maryland, College Park, MD
20742-2421, USA\\
$^{6}$Observatoire de Paris, LESIA-UMR CNRS 8109, 5 place
Jules Janssen F-92195 Meudon cedex, France\\
}
\begin{document}

\date{Last version: 22 January 2011}

\pagerange{\pageref{firstpage}--\pageref{lastpage}} \pubyear{2002}

\maketitle

\label{firstpage}

\begin{abstract}
We present several lines of evidence based on different kinds of
observations to conclude that rotational fission has likely occurred for a
fraction of the known Trans-Neptunian Objects (TNOs). It is also likely that
a number of binary systems have formed from that process in the
trans-neptunian belt. We show that Haumea is a potential example of an
object that has suffered a rotational fission. Its current fast spin would
be a slight evolution of a primordial fast spin, rather than the result of a
catastrophic collision, because the percentage of objects rotating faster
than 4 hours would not be small in a maxwellian distribution of spin rates
that fits the current TNO rotation database. On the other hand, the specific
total angular momentum of Haumea and its satellites falls close to that of
the high size ratio asteroid binaries, which are thought to be the result of
rotational fissions or mass shedding. We also present N-body simulations of
rotational fissions applied to the case of Haumea, which show that this
process is feasible, might have generated satellites, and might have even
created a ``family'' of bodies orbitally associated to Haumea. The orbitally
associated bodies may come from the direct ejection of fragments according
to our simulations, or through the evolution of a proto-satellite formed
during the fission event. Also, the disruption of an escaped fragment after
the fission might create the orbitally related bodies. If any of those
mechanisms are correct, other rotational fission families may be detectable
in the trans-neptunian belt in the future, and perhaps even TNO pairs might
be found (pairs of bodies sharing very similar orbital elements, but not
bound together).

\end{abstract}


\section{Introduction}

Our solar system contains a large number of icy bodies beyond Neptune's
orbit. These objects are collectively referred to as Trans-Neptunian objects
(TNOs), although they are also known as Edgeworth-Kuiper Belt Objects
(EKBOs), or simply, Kuiper Belt Objects (KBOs). These icy bodies are thought
to be leftovers from the process of solar system formation and are believed
to contain the most pristine material of the solar system beyond the ice
line. They are also thought to be the source of the short period comets
\citep{Fernandez1980}, although many details of the mechanisms that bring
the material from the trans-neptunian region to the inner solar system are
still missing. A wealth of knowledge on the trans-neptunian region has been
accumulating since the discovery  \citep{Jewitt1993}  of the first TNO
in 1992 (after Pluto and Charon), but the study of TNOs is still a young
field and there are still many open questions.

A topic that has attracted particular interest within TNO science is
binarity. Binaries are a powerful means to study the trans-neptunian belt
because they may allow us to derive masses and densities of their components
(by assuming some mean albedo value). Also, TNO binaries appear to be quite
common \citep{Noll2008}.

Several mechanisms of binary formation have been proposed for TNOs. Most of
them have been reviewed in \cite{Noll2008} and there are also newer binary
formation scenarios such as direct collapse \citep{Nesvorny2010}. However,
rotational fission has not been particularly investigated in the case of
TNOs. This mechanism is thought to be an important source of binaries in the
Near Earth Asteroid population of objects \citep[e.g.][]{Walsh2008}, whose
sizes and compositions are apparently very different from those of the much
larger TNOs that we can currently observe. Although the preferred formation
mechanisms of most of the binaries in the trans-neptunian belt is the
capture scenario \citep[e.g.][]{Noll2008}, rotational fissions might also
provide a fraction of the observed high mass ratio binary systems, and
other binaries with small specific angular momentum. It would be useful to
know approximately what fraction should be expected. The study of rotational
fission is not only important for binarity studies, but also for our general
understanding of the trans-neptunian belt.
\newline
In this paper we present some evidence to show that rotational fission of
TNOs is a relevant mechanism, especially for large TNOs and we study
Haumea's case in detail. Haumea (previously known as 2003 EL$_{61}$) is a
good candidate to study because of its large size and fast spin 
\citep{Rabinowitz2006}. We also present numerical simulations of
spontaneous rotational fissions of large TNOs, which we apply to Haumea's
case. In addition, we consider whether sub-catastrophic collisions can
induce the rotational breakup of primordial bodies that were already fast
rotators, and we discuss the stability of the binary/multiple systems formed
after rotational fissions.

\section{Observational clues for the existence of rotationally fissioned
bodies}

After studying the rotational parameters of several TNOs,
\cite{Ortiz2003} showed that a material strength of $\sim$1000kPa is needed
for the TNOs to withstand shear fracturing and remain intact. Therefore,
objects having a smaller material strength than that value would not be
intact: they would be damaged and would have fractures. We suspect that most
TNOs have smaller material strength than 1000kPa (because the material
strength of their relatives --the comets-- is orders of magnitude smaller
than this). Thus, we suspect that most of the TNOs would be structurally
damaged objects, that is partially or completely fractured bodies.
Therefore, at least some TNOs might be able to breakup easily due to
rotation. Besides, for ``large" TNOs, their mass would be sufficient to
overcome rigid body forces and therefore be in hydrostatic equilibrium. The
issue of how large these bodies must be in order to be in hydrostatic
equilibrium is still unclear \citep{Tancredi2008, Duffard2009} because there
are still a number of unknowns about the mechanical behaviour of the icy
mixtures that form the TNOs. For these kinds of bodies --not dominated by
rigid body forces-- the study of rotational fission from the perspective of
the physics of fluid bodies might be interesting.

From maxwellian distribution fits to the observed rotation rates of TNOs
\citep{Duffard2009}, we can immediately realize that the spin distribution
implies that $\sim$ 20$\%$ of very fast rotating objects would not be able
to remain in hydrostatic equilibrium for the typical densities of TNOs. Such
densities are likely around $1000$ to $1500~kg/m^3$. Figure 1 shows a
maxwellian distribution that fits the observed distribution of known
rotational periods of TNOs compiled in \cite{Duffard2009} with additional
data from \cite{Thirouin2011}.
A maxwellian distribution arises if the three components of the angular
velocity are distributed according to a gaussian with zero mean values and
equal dispersions; such distributions have frequently been compared to
histograms of rotation rates of asteroids \citep{Binzel1989}.

The spin frequency distribution we see today is the evolution of the
primordial one. The primordial spin distribution changed as a result of
frequent collisions in the early ages of the Kuiper Belt. At that epoch the
trans-neptunian belt was very massive and the collisional evolution was
intense \citep{Davis1997, Benavidez2009}. Because collisions can spin up or
spin down the bodies, the final distribution of rotations can include a
fraction of objects spinning faster than the average initial spin frequency.
We think that a fraction of the objects that underwent net spin-up ended up
suffering rotational fissions because they reached their critical rotation
speeds.

The shaded area in Figure 1 indicates the percentage of objects with a spin
faster than $\sim$4 hours that should be expected from our maxwellian fit.
Specifically, in \cite{Duffard2009} we show that $\sim$ 15$\%$ of the
objects cannot be equilibrium figures for a typical density of
$1500~kg/m^3$, whereas the percentage rises to 25$\%$ for a density of
$1000~kg/m^3$ \citep[Figure 6][]{Duffard2009}. In other words, around $\sim$
20$\%$ of the objects would have fissioned due to rotation.
Furthermore, there is additional observational evidence towards the
existence of a spin barrier around 3.9 to 4 hr in the observational data
\citep[e.g.][]{Thirouin2010, Duffard2009} below which no TNOs are found.
This possibly indicates that the bodies predicted in the maxwellian
distribution below $\sim 4\ hr$ have already broken up.

It can be argued that we do not see objects spinning faster than $\sim$ 4
hours simply because they could not form in the accretion phase. However,
our view is that those objects did not form, but the objects that formed
from the accretion phase suffered an intense collisional environment that
accelerated some of them and slowed down some others. Those TNOs that
suffered spin-up to a significant degree would undergo a significant
mass loss if their critical rotation periods were reached. As already
stated, we indeed know that there was an intense collisional evolution in
the early phases of the Kuiper belt and thus we think that the spins were
significantly altered in this phase. From this point of view, most of the
rotational fissions would have taken place in the first gigayear after the
formation of the solar system, when collisions were more frequent.

After a fission, at least part of the material ejected from the parent body
can form a satellite. In the case of asteroids, it is well known that the
formation of a satellite is one of the outcomes of rotational fission.
Similarly, binary or multiple systems might be or might have been common
within the trans-neptunian region. Nevertheless, if they are as old as a few
gigayear, the effects of dynamical interactions and subtle collisions could
have destroyed a large fraction of binary and multiple systems.

Since our paper \citep{Ortiz2003} we were expecting to find fast
rotators in the TNO population that would allow us to study these mechanisms
in detail. Haumea (formerly known as (136108) 2003 EL$_{61}$) turned out to
be an excellent candidate for that. Its very fast rotation
\citep[e.g.][]{Rabinowitz2006} could perhaps make it a typical case of a
rotational fission, and the existence of small satellites would also argue
in favour of the object being the remnant of a rotational fission process.
Thus, we chose this object as the best study case.

There are other observations that might indicate the existence of TNO
binaries originating from rotational breakup. One of these cases may be
Orcus' system. The specific total angular momentum of the system is very
close to that of an object with the same size and mass but spinning near its
critical spin rate. The details of the study for the case of Orcus and other
useful data are presented in \cite{Ortiz2011}. In the Near Earth Asteroid
(NEA) population a similar argument was made to point out that the mechanism
of rotational disruption appears to be the formation scenario for many
binaries \citep{Pravec2006}.

\begin{figure*}
\begin{center}
\includegraphics[width=15cm, angle=0]{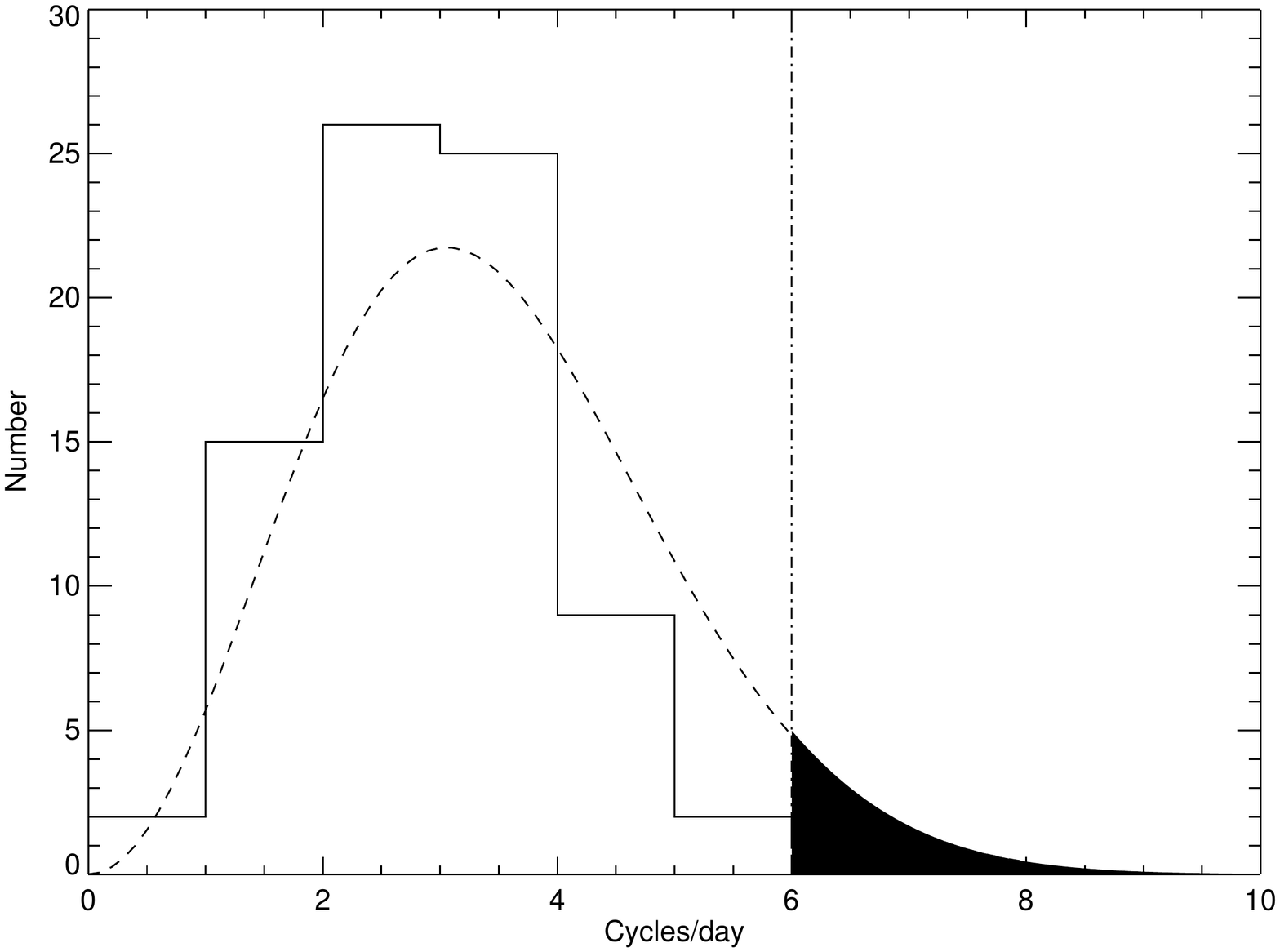}
\caption{Maxwellian distribution that fits the observational database on
rotation rates, taken from our work \citep{Duffard2009} plus recent results
\citep{Thirouin2011}. The black shaded area under the curve indicates the
percentage of objects that should spin faster than 4 h (6 cycles/day). Such
an area is not a very small fraction of the total area.}
\end{center}
\end{figure*}

\section{The case of Haumea}

2003 EL$_{61}$ (Haumea) is a dwarf planet with a tri-axial shape
($2000\times 1500\times 1000\ $km), a mass of $4.006\times 10^{21}\ kg$
\citep{Ragozzine2009} and a short spin period of $3.92\ hr$. Two satellites,
Hi'iaka and Namaka, are orbiting Haumea at 49880$\pm$198~km and
25657$\pm91$~km and have mass ratios relative to Haumea of $~1/200$ and
$\sim 1/2000$ respectively \citep{Ragozzine2009}. A group of TNOs has been
dynamically associated to this system and is frequently called Haumea's
``family''.
This has been imported from the asteroid belt, where it refers to groups of
objects that are very close in the proper elements space and comply with
suitable tests to establish their clustering.

It has been hypothesized that Haumea is a fast spinning object as a result
of a catastrophic collision that would have spun up the body and would
have --at the same time-- also created its two satellites and a collisional
family \citep{Brown2007}.
However, the claim that a catastrophic collision would have resulted in a
large body spinning quickly and, by serendipity, near its rotational breakup
limit is not supported by analytical or numerical works.
In fact, there is evidence to the contrary. \cite{Takeda2009} studied the
rotation end state of rubble-pile asteroids after collisions of different
sorts, by means of N-body numerical simulations, and showed that after
catastrophic collisions in a wide range of geometries, the largest remaining
body always rotated slower than prior to catastrophic collisions.

If these results for rubble piles are applicable to bodies in hydrostatic
equilibrium, the fast rotation rate of Haumea would not appear to be the
result of a catastrophic collision. It would be difficult to imagine that
Haumea had ever been rotating faster than today. In fact, the required
density and material strength --in the fluid approximation-- would have to
be even higher than its highest estimated density of around $2700\ kg/m^3$
\citep{Rabinowitz2006}, a much higher density than that of Pluto. Therefore,
it seems more plausible that Haumea was a fast spinning object from its
formation process.

On the other hand, using the collisional and dynamical evolution model
by \cite{Campo2011}, the probability of a catastrophic collision for a very
large object like Haumea is less than 7 $\times$ 10$^{-6}$
(\cite{Campo2011}, hereafter CB2011).

 One has to come up with very artificial mechanisms such as the collision of
two scattered disc objects, resulting in a classical belt one, to get a
small chance of a catastrophic event \citep{Levison2008a}.  Besides, the
alleged collisional ``family" of Haumea has estimated dispersion velocities
that are not consistent with the ones implied by the proposed collision.

Another collisional scenario has been put forward to explain Haumea's
``family'' by \cite{Schlichting2009}, who propose the formation of a large
satellite in an initial sub-sonic speed impact. The satellite would
subsequently be destroyed by means of a second collision and this process
would form the current two satellites together with the ``family'' itself.
Uncertainties in the collisional physics at sub-sonic speeds for objects
thousands of km in size and low probabilities ($<$ 0.3\%) for the
overall process to take place (CB2011) are potential weaknesses of this
model. Finally, in the time span required for the second collision, the
tidal interaction between the former satellite and Haumea would have slowed
down Haumea's spin so that its current fast rotation would not be explained.

The grazing collision scenario described in \cite{Leinhardt2010} has a
probability to happen less than 0.01\% after the Late Heavy Bombardment
(LHB) period and of less than 0.1\% before its end (CB2011). This scenario
also has trouble explaining the survival of the ``family'' after the onset
of the LHB phase some 4 Gyrs ago (CB2011), like the other scenarios. In this
phase -according to the Nice model
\citep{Gomes2005,Tsiganis2005,Morbidelli2005} - the mass of the region was
reduced to at most 5\% of the starting mass by dynamical effects. That
means that  the current mass of the family should have been at least 20
times larger, implying a larger parent body and an even more unlikely event
to create the system. The stability of the satellites in that phase is not
clearly granted either.

Because all the proposed scenarios meet difficulties, it seems natural to
explore a different scenario to explain Haumea's remarkable properties.
Rotational fission appears as a natural alternative process.
Here we propose that Haumea's parent body (which we call proto-Haumea) was
born already rotating fast
and subsequently suffered a rotational fission that perhaps created its
satellites and might have provided the mass of Haumea's ``family''.
In order to cause the spin up of an isolated rotating system, additional
angular momentum must be provided by an external cause.
In the Near Earth Asteroids case, a torque due to the YORP effect causes the
spin-up. We do not know the exact reasons for spin up in the trans-neptunian
region. Rotational fission may be induced by a sub-catastrophic collision
(those events were not unlikely at all, contrary to the catastrophic
collision scenario) providing enough angular momentum to trigger the
process. A moderately disruptive (non catastrophic) collision might have
transferred the slight amount of angular momentum needed to trigger a
rotational fission.
\cite{Takeda2007} have shown from numerical simulations that in moderately
disruptive impact events the largest remnant acquires a significant amount
of spin angular momentum. They stressed that in order for angular momentum
to be transferred to the spin of the largest fragment, the collision had to
be slightly disruptive, not catastrophic.

It is straightforward to show that for a generic triaxial ellipsoid with
size and mass close to those estimated for Haumea
rotating close to its critical angular momentum, a cratering collision with
a 300-500\ km size body
at typical Classical Disk relative velocities ($\leq$ 1\ km/s), off-axis
along the target's equatorial plane, would provide enough angular momentum
to trigger instability --and therefore mass loss-- on the proto-Haumea body.
This kind of collision was statistically relatively common ($\sim$ 1\%)
in the past, especially in the early Solar system up to the end of the LHB
phase, when hundreds to thousands of Pluto-sized objects still dwelled in
the disk.

As described in Section 2, from the maxwellian distribution that best fits
the current database on TNO rotations, we get that the percentage of objects
that should have ended up with rotation rates below $4$ hours is not small
(see Figure 1). Thus, we may expect that many TNOs acquired a ``high"
rotation rate.

\subsection{Haumea's satellites: Specific angular momentum}

The specific angular momenta ($H$) of the systems formed respectively by
Haumea + Namaka and Haumea + Hi'iaka are both around $0.3$ (see Figure 2),
while the scaled spin rate ($\Omega'$) is around $0.6$. We computed
 $H$ (Eqn. 1) according to \cite{Descamps2008} and
$\Omega'$  (Eqn. 5) according to \cite{Chandrasekhar1987}.
Specifically,

\begin{multline}
H =\frac{q}{(1+q)^{\frac{13}{6}}}\sqrt{\frac{a(1-e^{2})}{R_{p}}} +
\frac{2}{5} \frac{\lambda_{p}}{(1+q)^{\frac{5}{3}}} \Omega + \\
\ \frac{2}{5} \lambda_{s} \frac{q^{\frac{5}{3}}}{(1+q)^{\frac{7}{6}}}\left
(\frac{R_{p}}{a} \right)^{\frac{3}{2}}
\end{multline}
where {\it q} is the secondary-to-primary mass ratio, {\it a} the semimajor
axis, {\it e} the eccentricity, and {\it R$_{p}$} the primary radius.
The {\it $\Omega$} parameter is the normalized spin rate expressed as:
\begin{equation}
\Omega = \frac{\omega_{p}}{\omega_{c} }
\end{equation}
where $\omega_{p}$ is the primary rotation rate and $\omega_{c}$ the
critical spin rate for a spherical body:
\begin{equation}
\omega_{c} = \sqrt{\frac{GM_{p}}{{R_{p}^3}}};
\end{equation}
here {\it G} is the gravitational constant and {\it M$_{p}$} the mass of the
primary.
Assuming a triaxial primary with semi-axes as $ a_{o}> a_{1} > a_{2}$, the
$\lambda_{p}$ shape parameter is
\begin{equation}
\lambda_{p}= \frac{1+\beta^{2}}{2(\alpha\beta)^\frac{2}{3}}
\end{equation}
where $\alpha$ = $a_{2}/a_{0}$ and $\beta$ = $a_{1}/a_{0}$. \\
In this work, we considered the satellites to be spherical bodies, so
$\lambda_{s}$=1.\\
\newline
Finally,
\begin{equation}
\Omega'= \frac{\Omega}{\sqrt{\pi G \rho}}
\end{equation}
\newline
where $\rho$ is the density of the object.
\newline

The specific angular momenta and scaled spin rate of the systems formed by
Haumea + Hiiaka and Haumea + Namaka fall into the ``high size ratio
binaries" circle in Figure 1 of \cite{Descamps2008}. They studied the
binaries in the asteroid population (near Earth, main belt and Jupiter
trojan asteroids) and came to the conclusion that those systems very likely
arise from rotational fission or mass shedding. Therefore, Haumea's system
falls into that same class of binaries, which lends support to the idea that
the system may come from a fission process rather than a catastrophic
collision.

 \cite{Pravec2006} also pointed out that the specific angular momentum of
most asynchronous binary systems in the NEA population is similar (within
$20\%$ uncertainty) and close to the angular momentum of a sphere with the
same total mass (and density) rotating at the breakup limit. This suggested
to them that binaries were created by mechanisms related to rotation close
to the critical limit for break up.

In the next section, we turn to numerical simulations of the rotational
fission of a fast-spinning body gently spun up until it breaks up. We also
simulate a final rotational disruption triggered by a small impact.

\begin{figure*}
\includegraphics[width=17cm, angle=0]{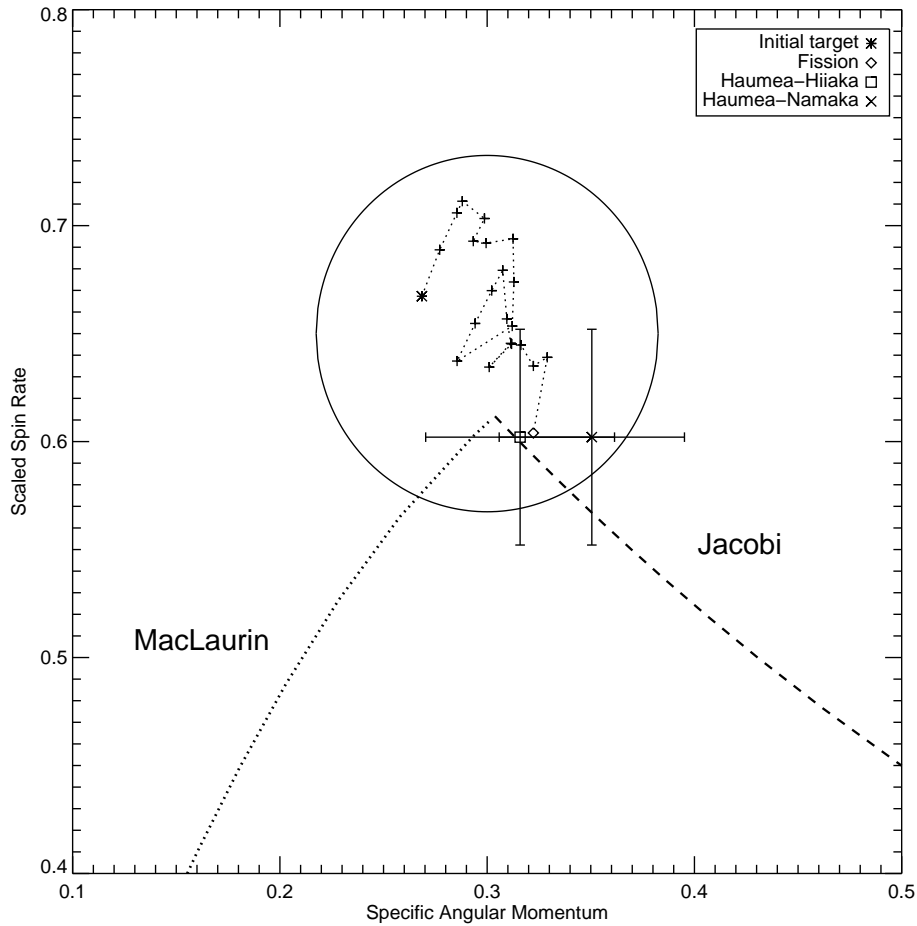}
\vspace{0.5cm}
\caption{Scaled Spin Rate versus Specific angular momentum of the systems
formed by Haumea-Namaka and Haumea-Hi'iaka (asterisk and square symbols,
respectively). Each cross represents a small increase of angular momentum in
a synthetic body described in section 3.2. The diamond symbol indicates the
point where the proto-Haumea underwent fission. The proto-Haumea fissioned
near the Jacobi-MacLaurin transition point in the high-size-ratio binaries
zone. The circle bounds the high-size-ratio binaries zone
\citep{Descamps2008}.}
\end{figure*}

\subsection{Numerical simulations of rotational fissions}

In order to carry out our fission simulations we have assumed that at
least a part of the TNOs are gravitational aggregates.
 \cite{Housen2009} performed laboratory experiments in which he shows that
$N$ collisions --each with energy $Q^*_S/N$, that is $1/N-th$ the threshold
specific energy for fragmentation of the target-- cause the same amount of
structural damage, into the target itself, as a single collision at $Q^*_S$.
Therefore, $N$ sub-catastrophic collisions can finally shatter a large
target without ejecting mass and producing a cohesionless structure that is
similar in many respects to a gravitational aggregate.

A gravitational aggregate behaves almost like a fluid when it comes to
rotation. The situation is not exactly the same  due to the presence of some
shear strength \citep{Holsapple2008} and it can be numerically handled with
the help of a suitable N-body code \citep{Tanga2009}. Therefore by studying
the rotational fission of gravitational aggregates we can also get an
approximation to the behaviour of rotating objects in hydrostatic
equilibrium, which, by definition, are dwarf planets. In other words, we do
not expect that TNOs larger than 1000 km are gravitational aggregates, as
their interiors are very likely in hydrostatic equilibrium, but the shape
they adopt and their general response to rotation can be approximated with
the gravitational aggregate structure.

With the aim of studying the possibility of forming binary systems by
rotationally fissioning large gravitational aggregates, we performed
numerical simulations of the processes using the PKDGRAV N-body code
\citep{Richardson2000, Stadel2001, Richardson2009}.
This code has the advantage of performing both the numerical integration of
mutual gravitational interactions between the mass components (considered as
hard spheres) of a given gravitational aggregate, and the calculation of the
collisional interactions between any pair of such components.
Gravitational aggregates have shear strength owing to the finite particle
sizes and the confining pressure of gravity. This is automatically accounted
for by PKDGRAV. On the other hand, shear stress (resistance to sliding) is
not included in the code, but instantaneous rotation of components is
considered whenever a collision occurs. It is straightforward to show that
the work necessary to move a cubic mass across one of the faces of an equal
mass cube, in the presence of friction, is only $28\%$ larger than the work
necessary to rotate a sphere --with the same volume as the cube's-- over
$1/4-th$ of the surface of an equal mass sphere. So, the code is
underestimating surface friction in this case. Nevertheless, if the
calculation is made considering cubes and spheres with equal surfaces
(instead of volumes), the equivalent work is $17\%$ smaller in the case of
the sliding cubes than in the case of the rotating spheres, and now the code
is overestimating surface friction. In any case, as the true situation
inside a gravitational body involves both dissipative sliding and rotation
of irregularly shaped components, and as the two calculated effects are of
the same order, it can be assumed that the code accounts for surface
friction to some extent.

Coming to the numerical simulations that have been performed, the first step
of the process is the generation of a fast spinning object with a total mass
around $4.5\times 10^{21}\ kg$.
 Such a gravitationally held object has comparable mass and size to Haumea,
with some $10\%$ larger mass to account for mass loss as the system is
formed. The proto-Haumea body is generated by means of a coagulation method
starting from a spinning nebula of $1000$ equal-sized particles that
generates a stochastic pile of spheres with no preferential geometrical
structure \citep{Tanga2009}. The physical characteristics of a typical
proto-Haumea generated in this way are listed in Table 1.

The scenarios mentioned in the previous section for the formation of a
primary and a satellite were studied by means of 4 sets of simulations:

S1) Sequences of gentle spin-ups of the parent body. 21 small increments of
angular momentum were performed until fission occurred.
The object is allowed enough time to adjust itself to the corresponding
equilibrium figure of rotation between successive angular momentum
increments.
This technique is used in order to look for the object's disruption limit in
a very smooth way, avoiding sharp accelerations to the body's rotation.

S2) These simulations are induced rotational fissions. They are equivalent
to the above scenario until the twentieth spin-up step is done.
This was done to simulate a situation in which a proto-Haumea is originally
rotating fast and at some point a low-energy collisional event happens.
The last step is performed by means of a  collision that provides enough
angular momentum to trigger fission. The relative speed of the collision is
1~km/s, the average impact speed in most of the Main Classical Belt of
TNOs.
This simulation is performed in order to answer to the straightforward
question that may arise after S1: why should a 2000\ km--size body increase
its own angular momentum at some point? In the asteroid belt, the YORP
effect is able to spin up bodies up to a few km in size, and close
encounters with planets may have a similar effect on NEAs too. Nevertheless,
no effect like YORP is available for a body of Haumea's size and at
heliocentric distances on the order of 40 AU, nor are planetary close
encounters likely in the trans-neptunian region. Comets can speed up their
rotations from the torques created by sublimating material on their
surfaces. However, this effect will also be too small on TNOs, which are
considerably larger than usual comets. The most likely process capable of
triggering the fission of an already fast-spinning TNO seems to be a
collision.

S3) A faster collision than in scenario S2, which provides more angular
momentum than strictly needed for a fission. The collision is performed at
3~km/s. The relative speeds that have been tested are close -or even above-
the limit for sound speed in the target body. In a homogeneous body,
simulations of hyper-velocity collisions must include the damage produced by
the propagation of the shock wave into the body structure, as is done in
Smoothed-Particle Hydrodynamics (SPH) simulations. Nevertheless, this
consideration does not invalidate our technique because we are dealing with
bodies that have -at least- a crust of heavily fragmented material. In such
environments, the shock wave is rapidly extinguished \citep{Asphaug1999}:
the damage is limited to the collisional area where part of the energy is
dissipated and the rest of the energy is available for dissipative
collisions and rotations to occur between the fragments forming the
outer structure of the body itself.

S4) This fourth scenario corresponds to simulations in which a different target 
is impacted by the projectile at 3~km/s. Except for the target, this scenario
is the same as S3. In S4 the target has a different number of particles and 
rotation period compared to S2 and S3. The characteristics of this target and those of scenarios 
S2 and S3 are listed in table 1.

\begin{table*}
\caption{\label{tab1} Physical characteristics of Target 1 (the proto-Haumea
generated for the simulations of the pure rotational fission scenario, S1).
Target 2 is the body created after the 20-th spin up of scenario S1. Target
2 is used in the collisionally induced rotational fission (scenarios S2,
S3). Target 3 is the target used for the S4 scenario. 
Also listed are the physical properties of the projectile used for the
simulations S2, S3 and S4. N is the number of particles; $a_{1},a_{2},a_{3}$ are
the semiaxes of the body; $\rho_{b}$ is the initial bulk density; $T_0$ is
the initial rotation period.}
\begin{tabular}{@{}lccccc}
\hline
 Object   & N & Mass [kg] & a$_{1}$, a$_{2}$, a$_{3}$ [km] &  $\rho_{b}$
[g/cm$^{3}$] & T$_{0}$ [hr]  \\
\hline
Target 1  & 866 & $4.48\times 10^{21} $ & $1362\times 744\times 513$ & 2.1 &
3.98   \\
Target 2   & 797 & $4.12\times 10^{21} $ & $1620\times 611\times 483$ & 2.1
& 4.52   \\
Target 3   & 846 & $4.38\times 10^{21} $ & $1355\times 641\times 506$ & 2.4
& 3.64   \\
Projectile                  & 183  &  $1.92\times 10^{20} $ & $349\times
338\times 294$ & 1.3 & No rotation  \\
\hline
\end{tabular}

\end{table*}

\begin{table*}
\caption{\label{tab2}Some results of the simulations. $M_p$ and $M_e$ are
the masses of the primary and the mass ejected from the system,
respectively; $M_s/M_p$ is the mass ratio of the binary system (mass of the
satellite divided by mass of the primary); $T$ is the rotation period of the
primary;  $<V_d>$ is the average speed of ejected free particles with
respect to the centre of mass, of ejected pairs of particles, and of ejected
rubble piles, respectively.}
\begin{tabular}{@{}lccccc}
\hline
Simulation   & $M_p$ [$\times 10^{21}\ kg$] & $T$ [h] & $M_s/M_p$ & $M_e$
[$\times 10^{20}\ kg$] &  $<V_d>$ [m/s]  \\
\hline
S1    & 3.922 & 3.698 & 0.113 & 3.620 & 303, 429, 318 \\
S2    & 4.302 &  3.823  & 0.113 & 1.327 &  490, 0$^{*}$, 0$^{*}$\\
S3    & 3.460 &  3.375  & 0.237  & 0.576& 1296, 0$^{*}$, 0$^{*}$ \\
S4    & 4.160 &  3.632 & 0.017  & 3.398& 1912, 1009$, 330$ \\
\hline
\end{tabular}
\\ $^\mathrm{*}$ in these simulations, no groups of two particles and
rubble-piles were formed.
\end{table*}

 Dozens of simulations were performed within each of the three scenarios.
Fission easily results in the formation of a pair of objects with positive
total energy, or in the formation of a bound system (binary) in S1 and S2.
For any set of simulations, a representative sample of boundary conditions
is chosen here for description (Table 2).
In the case of S3, the production of a bound system is restricted to a
narrow range of boundary conditions.

In Figure 2 we plot the scaled spin rate versus the specific angular
momentum for the 21 steps of Simulation n$^\circ$1. As can be seen in the
plot, the proto-Haumea fissions near the Jacobi-MacLaurin transition point.
Animations showing the three fission scenarios are presented as online
material. In Figure 3 we show the speed distribution of the ejected material
in the three different scenarios. In all cases the fragments escaping the
system immediately after rotational fission have average speeds of 0.3~km/s,
0.5~km/s 1.3~ km/s and 1.9~km/s for scenarios S1, S2, S3 
 and S4 respectively.  However, the distribution of ejection speeds 
is very broad. See Figure 3. In Figure 4 we show snapshots of the simulations.

In many simulated cases a large-enough body is formed from the ejecta of the
parent body  and remains in orbit around the primary for the full length of
the numerical integration (several days). By a large-enough body we
mean an object with the total mass of the ``family" and the satellites. 
The stability of the binary systems formed has not been studied numerically
with PKDGRAV because the long-term evolution is very CPU intensive, but it
can be analyzed from theoretical and other methods (see Discussion section).
\begin{figure}
\includegraphics[width=7cm, angle=0]{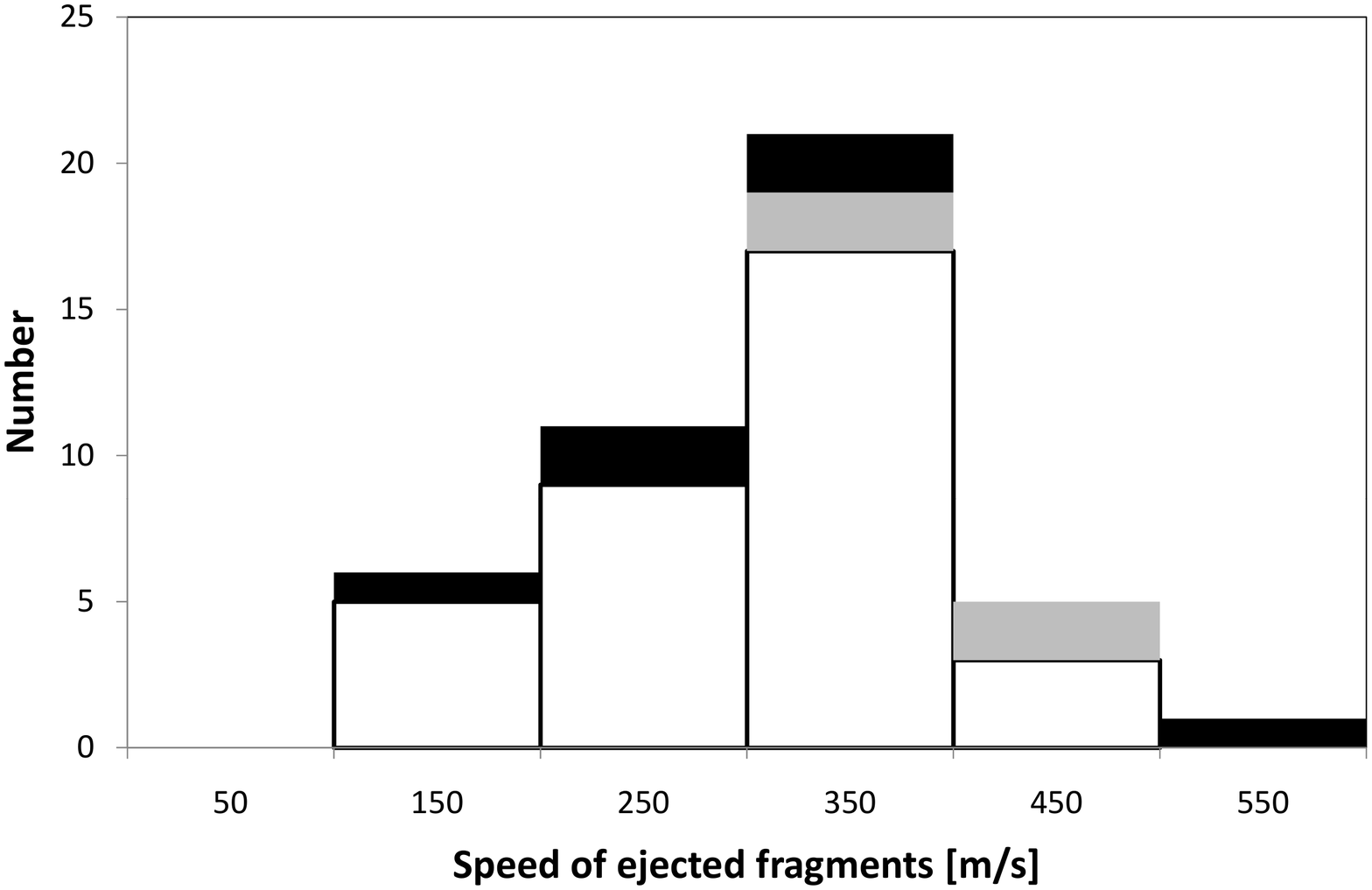}
\includegraphics[width=7cm, angle=0]{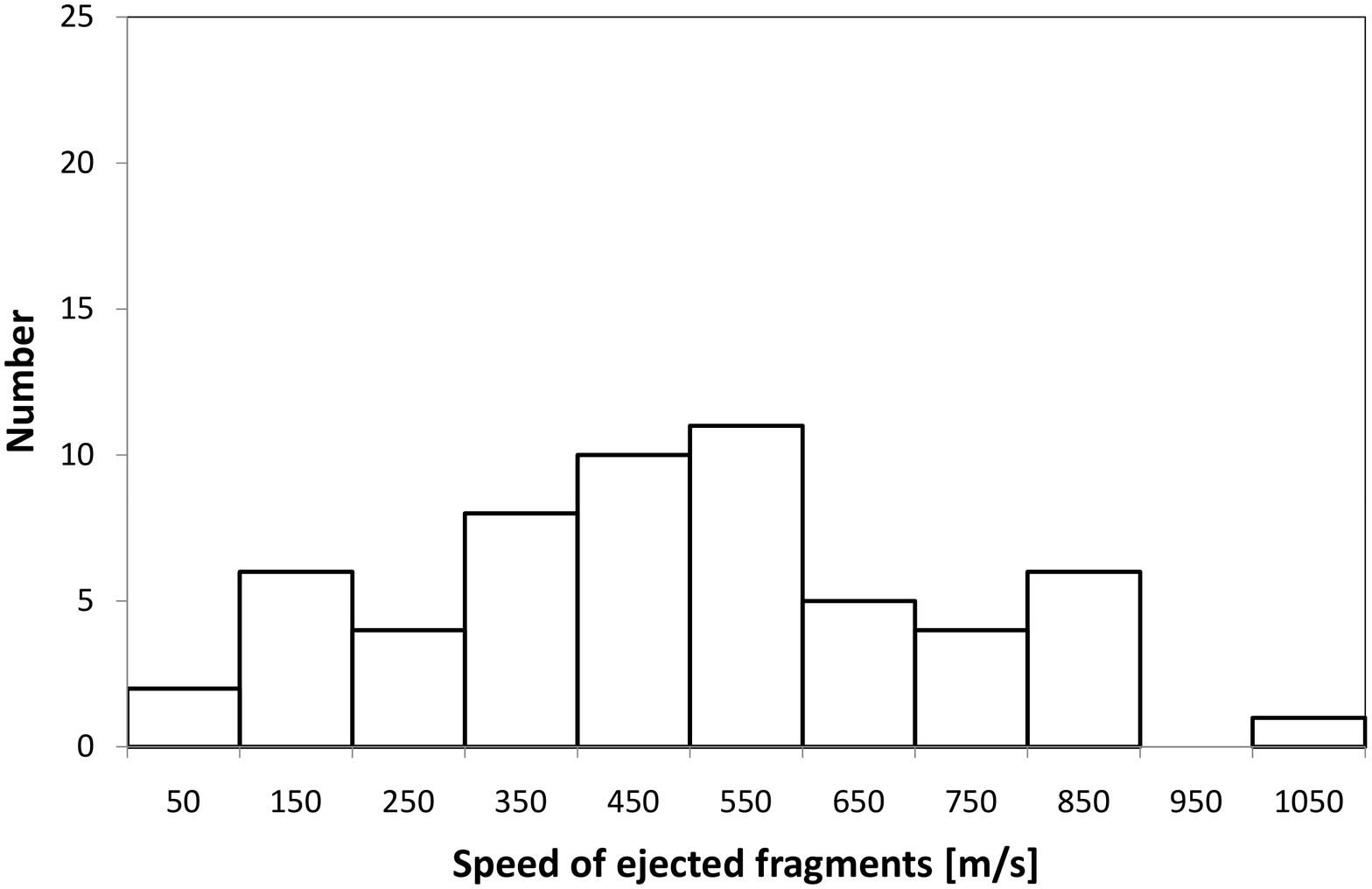}
\includegraphics[width=7cm, angle=0]{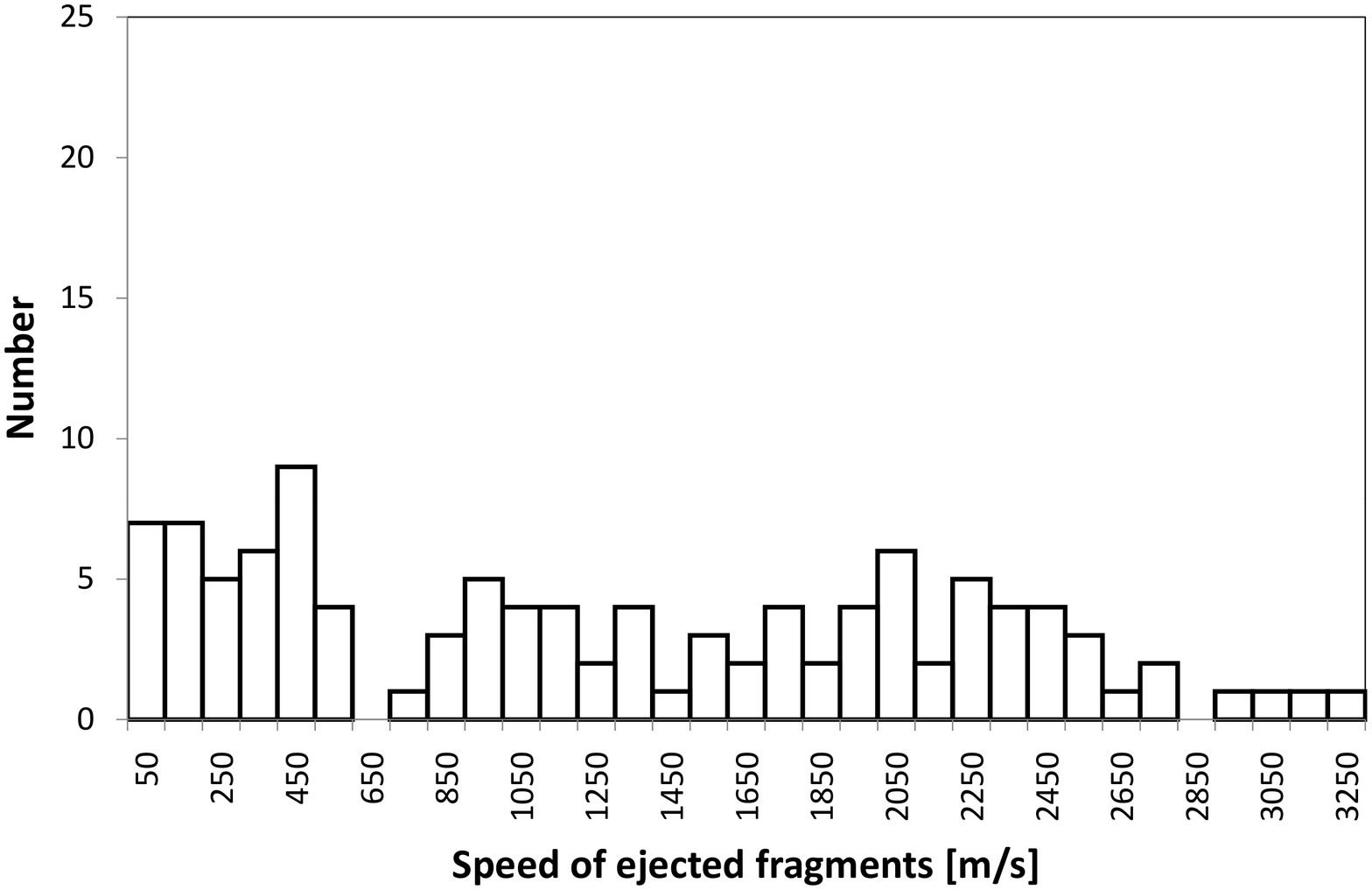}
\includegraphics[width=7cm, angle=0]{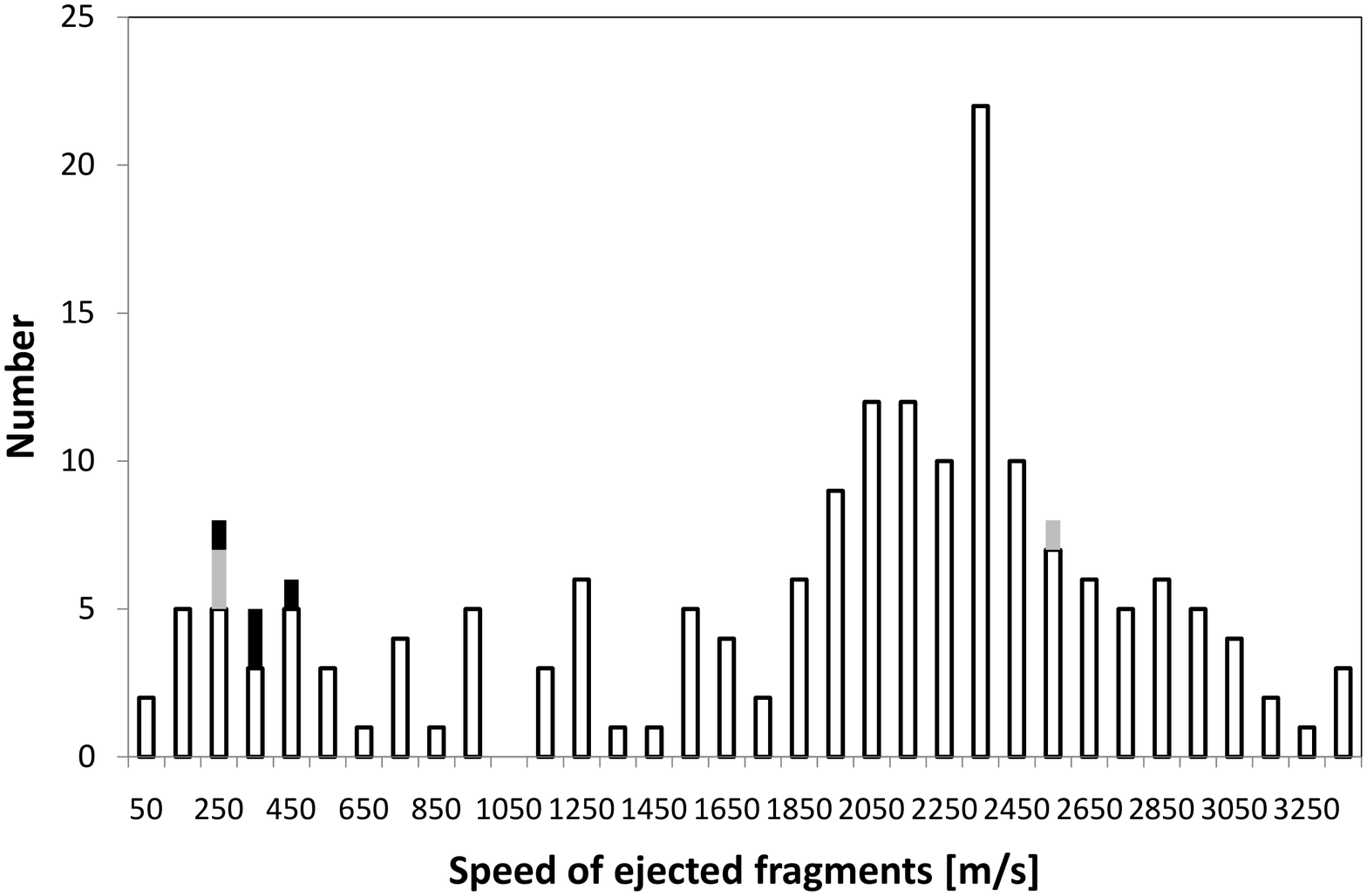}
\caption{ Histograms showing the number of ejected fragments as a function
of the speed with respect to the centre of mass in simulations S1 (upper
plot),  S2 (second plot), S3 (third plot) and S4 (bottom plot). The gray bars in the upper
plot correspond to groups of two particles and the black bars correspond to
ejected rubble piles, meaning a group of three or more particles.
Simulations S2 and S3 did not produce groups with two or more particles.
Movies of these simulations are presented as online material. S1 corresponds
to pure rotational fission whereas S2, S3 and S4 correspond to collisionally
induced rotational fission with impact speeds of $1000 ~m/s$, $3000 ~m/s$ and $3000 ~m/s$
respectively. See Section 3 of the text.}
\end{figure}

\section{Results and discussion}

From the reported numerical simulations and from other considerations  made
we suggest that the fission mechanism for the formation of large complex
systems of TNOs --like Haumea-- seems to be preferable (from a statistical
point of view), with respect to catastrophic collisions between large
primitive bodies.

It must be pointed out that using a very large gravitational aggregate to
describe Haumea is a considerable simplification because Haumea may be a
differentiated body \citep{McKinnon2008} with at least a fluid-like
interior. Nevertheless, gravitational aggregates behave almost like fluids
regarding the shape they adopt as a response to rotation. They even breakup
near the theoretical limit for a fluid (as shown in Figure 2). Hence, the
simulations presented in this paper retain the basic physics of rotational
fission even for large bodies like Haumea.

We now turn to examining whether the rotational fission mechanism alone can
reproduce all the observables of the Haumea system (subsection 4.1). In
subsections 4.2, 4.3 and 4.4 we speculate whether three other related
mechanisms may also explain the observables. The main observables are the
existence of satellites, the fast Haumea spin, and the existence of a
``family" with a 140 m/s dispersion speed.

\subsection{Rotational fission alone}

By rotational fission we mean any of the S1, S2, S3 and S4 scenarios mentioned
in the simulations section. That is, pure rotational fission
(S1) --regardless of its cause--  or collisionally induced rotational
fission (S2, S3, and S4).

Even though the numerical simulations form satellites of various sizes
together with a fast spinning Haumea, which are two of the main observables,
the formation of a family has to be explained as well. Looking at the
distribution of speeds (Figure 3 and Table 2) it would seem that the family
is not formed because the average ejection speeds in scenarios S1, S2, S3 and S4
are much higher than 140 m/s. On the other hand, it must be pointed out that
Haumea itself requires a dispersion speed of 400 m/s whereas the rest of the
members of the family cluster around a dispersion speed of 140 m/s
\citep{Brown2007}. Therefore, the fragments ejected from Haumea need an
offset speed of $\sim$ 300-500 m/s with respect to Haumea itself.

In this regard, let us point out that the ejected fragments in our
simulations have a net predominant direction: by taking the average of the
velocity vectors (at infinity with respect to the centre of mass of the
system) of all the ejected fragments, we come up with a vector of
components $(13\  m/s,\ 22\ m/s ,\ 0\ m/s)$ with a modulus of $25.2\ m/s$
and standard deviation of $328\ m/s$ in scenario S1. For scenario S2 we get
$(-447\ m/s,\ -189\ m/s,\ -34.5\ m/s)$ with a modulus of $487\ m/s$ and a
standard deviation of the speed around this direction of $314\ m/s$. For
scenario S3, the mean velocity vector is $(-934\ m/s,\ 442\ m/s,\ 200\ m/s)$
with a modulus of $1050\ m/s$ and standard deviation of $1250\ m/s$. For
scenario S4 we come up with $ (-1730\ m/s, \ 263\ m/s,\ 11\ m/s) $ and a 
modulus of $1750\ m/s$ with a standard deviation of $1131\ m/s $.

Because Haumea has an offset speed of $400\ m/s$ with respect to the rest of
the members of the ``family" \citep{Brown2007} and because this offset speed
can be reproduced with the S2 scenario, we think that this scenario is our
best approximation to explain the formation of the Haumea system. However,
the dispersion speed of $328\ m/s$ of the fragments is still a factor of 2.3
higher than needed.  One should note that the 400 m/s offset of Haumea
due to its displacement in eccentricity from the remainder of the family
might be explained by Haumea's chaotic diffusion within the 12:7 mean-motion
resonance with Neptune, which can change Haumea's eccentricity to its
current value \citep{Brown2007}. In our model this eccentricity difference
can be explained if the material is ejected in the orbital plane. In that
case the orbits of the ejected fragments will have a very different
eccentricity with respect to the progenitor, but not a significantly
different inclination. If the spin axis of the proto-Haumea was
perpendicular (or nearly perpendicular) to its orbital plane the ejection of
the fragments would be in the orbital plane, so we would expect a small
spread in inclinations and a larger separation in eccentricity with respect
to the progenitor. Thus we do not need to invoke chaotic resonance diffusion
to explain the different eccentricity of Haumea to the rest of the family
members.

In summary, scenario S2 is qualitatively consistent with the observables and
quantitatively very close to the exact values of the observables. A slightly
smaller impact speed below 1000 m/s might provide more precisely the offset
speed and the dispersion speed observed in the Haumea system. The
offset --with respect to the family members-- in Haumea's eccentricity and
not in inclination is a consequence of the fission happening close to the
orbital plane. The family members are part of the ejected components from
the parent body. This circumstance is likely because large bodies in many
cases have small obliquities.

Our simulations can form a large satellite (see Table 2) and the family, but
only in some cases --belonging to the simulation series S1 and S2-- is a
second small satellite obtained. In addition to this difficulty the
dynamical coherence of the family (its velocity dispersion) would have been
destroyed if the collision that induced the fission took place when the
Kuiper Belt was more massive.

Although the induced rotational fission is our preferred mechanism to
explain the main features of the Haumea system, in the next subsections we
explore other dynamical mechanisms that might also place the shed material
in heliocentric orbits sufficiently grouped in orbital parameter space to
form the ``family" and simultaneously meet the other observables.

\subsection{Rotational fission plus collision on the proto-satellite}

A catastrophic collision on a large proto-satellite (some 500\ km in
diameter) formed after the fission can be an alternative mechanism to
generate a ``family" with the observed dispersion speed. At the same time it
could also generate the satellites.
This collision would not require very large impacting bodies and would thus
be reasonably likely with respect to a high-speed giant collision into
Haumea's parent body. In fact, the size distribution of TNOs is steep in the
required size range ($N(D, D\pm dD)dN \propto D^{-b}dD$, with $b>4$) and the
probability of a shattering event on a 500\ km size body, within an even
rarefied (that is, post-LHB phase) Classical Disk, is at least $4$ orders of
magnitude larger than that of having a catastrophic collision between two
bodies of about 1000\ km each. Specific simulations consisting of impacting
the fissioned satellite and getting a system with the current
characteristics of the Haumea system are currently underway, but that study
is beyond the scope of the present work, which is focused on the fission
process itself.

This scenario meets similar problems to that proposed by
\cite{Schlichting2009} and pointed out in section 3. Besides, the time span
between the formation by fission and the required impact event may be enough
to slow down Haumea's rotation through the tidal interaction of the
satellite. Angular momentum conservation implies that the orbital
momentum gained by the satellite is obtained from the rotation of the
primary and therefore the primary must slow down. 
Since Haumea's rotation is still very fast, the scenario of a collision on
the proto-satellite requires an impact shortly after the fission event (so
that the tidal effect does not have time to slow down Haumea), which is less
likely.

\subsection{Rotational fission followed by secondary fission of the
proto-satellite}

On the other hand, \cite{Jacobson1} recently proposed that low-mass-ratio
binary asteroids resulting from fissions are generally unstable, but stable
cases arise when the satellite suffers spin-up through tidal interactions
with the primary and finally undergoes a rotational fission itself, with
dispersal of part of the mass of the system. Thus, the same mechanism might
be applicable to TNOs and explain the existence of a group of bodies with
orbital elements related to those of Haumea (with small dispersion speeds).
This scenario does not require collisions. If the mechanism of rotational
fission of the secondary mentioned in \cite{Jacobson1} is not rare,
rotational fission families might be found around other large TNOs.
\cite{Jacobson2} point out that the spin up of the satellite and its fission
can only take place in systems with satellite-to-primary mass ratios smaller
than 0.2.

Therefore, if the formation of the Haumea ``family'' was the result of a
secondary fission (fission of the proto-satellite), the mass of the
``family" and the current satellites must be smaller than $0.2$ times that
of Haumea. This appears to be the case. In fact, summing up the mass of all
the members of the ``family" --computed by assuming an average albedo of
$0.6$ and a density of $2000\ kg/m^3$-- we end up with a mass that is just a
few percent that of Haumea, on the same order of the mass of the known
satellites. The uncertainty in mass comes primarily from the albedo
uncertainty, but --since all the objects clearly contain water ice in large
amounts-- they are believed to be at least as reflective as Haumea, so
albedos even higher than $0.6$ would apply. Recent and accurate measurements
of the albedo of one of the ``family'' members resulted in a value of
$0.88^{+0.15}_{-0.06}$ according to \cite{Elliot2010}. Therefore, the total
mass of the family may be even smaller than a few percent that of a Haumea.
The low mass ratio is a further clue in favour of the fission mechanism.

\subsection{Rotational fission, formation of a pair and disruption of the
small member of the pair}

An interesting mechanism for the formation of TNO systems arises as a
by-product of our numerical simulations of the Haumea system.
In some cases, the proto-Haumea fission results in the formation of a TNO
pair, with a secondary typically on the order of some 200-500\ km.

Referring to the Near Earth population and the Main Asteroid  Belt,  the
existence of asteroid pairs (pairs of asteroids with similar orbits but not
bound together) \citep{Vokrouhlicky2008} has been explained to arise from
rotational fissions. By means of dynamical simulations of the evolution of
fissioned bodies, \cite{Jacobson2} pointed out that systems with
satellite-to-primary mass ratios larger than $0.2$ always evolve to
synchronous binaries, whereas asynchronous binaries, multiple systems and
asteroid pairs can only form if their mass ratios are smaller than $0.2$.
\cite{Pravec2010} showed --from a large observational data set-- that
asteroid pairs are indeed formed by the rotational fission of a parent
contact binary into a proto-binary, which subsequently disrupts under its
own internal dynamics soon after formation. This is found only for mass
ratios smaller than $0.2$, as expected from the theory. These results
together with our numerical simulations suggest that pairs may have been
formed in the trans-neptunian region.

It has been shown that the primaries of the asteroid pairs have larger
lightcurve amplitudes than the primaries of binary asteroids with similar
mass ratios. This probably indicates that the elongated shapes of primaries
play a significant role in destabilizing the system and ejecting the
satellite \citep{Pravec2010}. The formation of a pair in the case of systems
with a primary with the characteristics of Haumea then seems plausible.

According to \cite{Jacobson2}, the time span in which a binary system ejects
its satellite is usually very short, therefore the tidal interaction would
not slow down the primary significantly and it may still be observed in a
high rotation state.

Once a TNO pair is formed, the secondary may subsequently undergo a
disruptive collision or a secondary rotational fission, so that a group of
bodies could be created. These objects would share very similar orbital
parameters to those of the primary and they would look like its collisional
``family''. Actually, the secondary would be their parent body rather than
the primary itself. The velocity dispersion of the fragments ejected in the
collision would indeed be close to the typical escape speeds from a 500 km
size  body, as happens in the case of the Haumea ``family" (140 m/s).
According to our simulations of spontaneous or induced fissions, most of the
fragments that escape shortly after have relative speeds with respect to the
primary around $400-500\ m/s$, in the range of the offset speed of Haumea
with respect to the rest of the ``family" members ($\sim$ 400 m/s). Figure 3
shows many fragments with escaping speeds in the required range.

A disruptive collision on a small object (the secondary of the pair) is
likely enough, so this  scenario is plausible to explain a group of bodies
with similar orbital parameters to that of the primary, as happens in the
case of Haumea.
Nevertheless, this ``family" formation scenario faces some difficulties in
the case of the Haumea system. In fact, the existence of two satellites
would not be straightforward to explain, but would not be impossible. For
example, a multiple or triple system might form soon after  fission, so that
the system ejected one of its satellites and retained the currently observed
couple of Haumea's satellites. Simulations of the S1  series show that this
is possible. \cite{Jacobson1} pointed out that a fraction of low-mass-ratio
proto-binaries can evolve to multiple systems that may eject one of its
members.

Also, the interaction of a third body with the proto-binary formed in the
fission process
might result in the ejection of the proto-satellite from the system at a
small relative speed with respect to Haumea. In this case, the mass ratio
would not have to be smaller than $0.2$. As explained above, if the ejected
body underwent a catastrophic disruption, the generated fragments would
likely share similar orbital parameters to Haumea's.
The interactions of binaries with third bodies were studied by
\cite{Petit2004} to estimate the stability and persistence of the primordial
binaries. They found that these interactions were frequent early on in the
solar system and a large fraction of binaries were destroyed. Therefore,
such a mechanism might also have taken place for a young binary Haumea.

In summary, a mechanism that might account for all the observables would
require that the  proto-Haumea fissioned and formed a stable low-mass-ratio
triple system (which is one of the outcomes of the evolution of rotational
fission of proto-binaries within the Jacobson and Scheeres (2010) formalism)
that would explain the presence of  the satellites Hi'iaka and Namaka. At
the same time, part of the ejected mass should have the correct dispersion
velocity to form the observed ``family'' or be clustered into a single
escaping body that should ultimately undergo a catastrophic disruption,
forming the ``family'' itself.

\begin{figure*}
\includegraphics[width=20cm, angle=0]{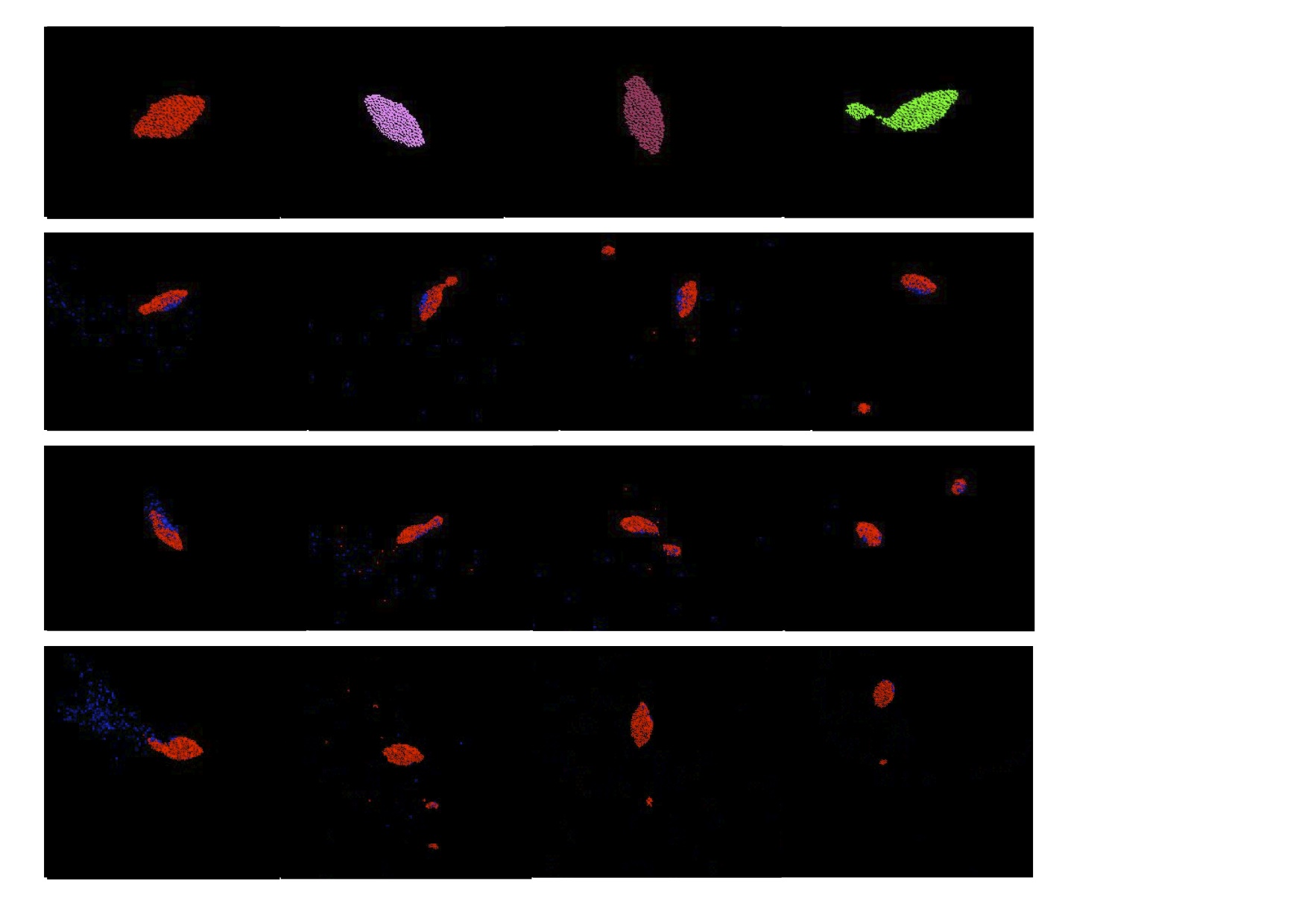}
\vspace{0.5cm}
\caption{ Snapshots of the different simulations. Top row is the
rotational fission scenario showing differents steps of the process in
Figure 2. Different colors were used every time the object was spun-up (from
left to right), following the same color coding as in the onlinefission1.avi
movie. Middle row is the S2 scenario showing the collision at 1 km/s. As can
be seen, part of the projectile material ends up in the crust of the large
body, covering a non-negligible area of the body. This might perhaps account
for the existence of a dark albedo area on Haumea \citep{Lacerda2008}. Third
row is the S3 scenario showing the collision at 3 km/s. Lower row is the
S4 scenario, in which a lower mass satellite is created compared to S3.
In the S2, S3 and S4 scenarios the projectile particles are depicted in blue.  }
\end{figure*}

\subsection {Future prospects}

 Can we find more ``families" similar to that of Haumea for other known
objects? Observationally we should try to find fast-rotating ``large" TNOs
with large lightcurve amplitudes and then look for objects with similar
orbital elements. However, among  potentially large TNOs, the only other
fast-spinning object is (120178) 2003 OP$_{32}$ \citep{Thirouin2010,
Rabinowitz2008}, which belongs to Haumea's ``family'' itself. Other fast
rotators in the $\sim 4\ hr$ period range could  be identified in the
future. Nevertheless, the tidal interaction of a former satellite may have
slowed down the spin of potentially good candidates and therefore current
very fast spins might not necessarily be a constraint. Varuna, the most
elongated object among the large TNOs, might be an interesting case of a
primary slightly slowed down. Unfortunately, there is currently no
indication that it has orbitally related objects.  If a ``family'' were
related to Varuna (magV $\sim$ 20), the members would be at least 2 to 3
magnitudes fainter than Varuna itself and the census of TNOs down to
magnitude $23$ is far from complete.

It may be the case that only the Haumea system has been found because it is
one of the brightest TNOs and its ``family'' members were detectable by
telescopic surveys.

A possible test for the relevance of the proposed fission mechanisms may be
derived by considering the resulting binary fraction. If all the
rotationally disrupted objects formed stable satellite systems after
rotational breakup, we should expect on the order of $20\%$ binaries for a
nominal bulk density of $1300\ kg/m^3$, as  discussed in Section 2. The
fraction of stable binary systems may be considerably lower than $50\%$,
because their stability depends critically on the mass ratio of the system
\citep{Jacobson2}. Keeping in mind that mass ratios larger than $0.2$ form
stable systems \citep{Jacobson2}, an average of $\sim 50\%$ of the fissioned
bodies might be stable and most of them should already be synchronous
binaries. Thus, within the large TNOs, around $10\%$ of them could be
binaries formed by rotational fission. This fraction may be lowered down if
third-body interactions occurred frequently in the young trans-neptunian
belt. Collisional evolution models that take into account changes in
rotation rates and are able to keep track of the surviving binary systems
would be needed to assess the fraction of binaries currently expected.

Concerning the possibility of finding ``TNO pairs", note that the orbital
elements of most TNOs are more uncertain than those of main belt asteroids.
Therefore, searches for TNO pairs are more difficult. Moreover, there are
only around $1400$ known TNOs. This is too small a sample if compared to the
around $5\times 10^5$ known asteroids, among which only $\sim 60$ pairs were
found \citep{Vokrouhlicky2008}. Besides, the small mass ratio implies that
many TNO pairs may remain undetected because one of the members is too
faint. Another difficulty resides in the fact that a large fraction of the
pairs might have formed a few gigayears ago and therefore they would be more
difficult to identify than in the asteroid belt, where pairs are much
younger than 1 gigayear. Nevertheless, we can perhaps bound the search
following \cite{Pravec2010}, who showed that the primaries of asteroid pairs
have larger lightcurve amplitudes than the primaries of similar-mass-ratio
binary systems. If this were applicable to TNOs, the best candidates to be
the primaries of TNO pairs are those with high lightcurve amplitudes, such as
Varuna, Haumea and a few others.


\section{Summary and Conclusions}

We have presented evidence indicating that rotational fission of TNOs may be
a mechanism that has affected a fraction of the TNO population. Binaries may
have formed that way in the trans-neptunian region. Also ``TNO pairs" might
exist as a result of rotational fission, and even triple systems. Binaries,
pairs and triple systems are the typical outcomes of rotational fission in
the asteroid population, depending on the mass ratio of the proto-binaries
formed. The indications for fissions in the trans-neptunian region come from
various observations and also from numerical simulations of the process.
Haumea is a particularly good candidate that might have suffered a
rotational fission because of its fast spin rate and other remarkable
features. Haumea's satellites might have been formed as a result of the
fission itself. The ``family'' of bodies orbitally related to Haumea may
derive from the ejected fragments after the fission (as we showed with our
simulations of type S1), or may also be a result of the evolution of a
proto-satellite in the proto-binary after the fission, or might even arise
from the disruption of an escaped fragment or an escaped satellite. We show
that the fission process has a larger probability of occurring than the
high-energy collisional scenarios that have been proposed in the literature
to explain the existence of satellites and bodies orbitally related to
Haumea. Therefore, we propose that the fission mechanism is a more natural
scenario and may generally explain most of the features of the Haumea
system. Future studies of high-mass-ratio binaries in the trans-neptunian
belt may provide more detail on the rotational fission scenario. Also, the
existence of ``TNO pairs", or future discoveries of other groups of
dynamically related objects, may shed more light on rotational fission of
TNOs.

\section{Online material}

Online material is available at 

\begin{verbatim}
www.iaa.es/~ortiz/onlinefission1.avi
\end{verbatim}

\begin{verbatim}
www.iaa.es/~ortiz/onlinefission2.avi
\end{verbatim}

\begin{verbatim}
www.iaa.es/~ortiz/onlinefission3.avi
\end{verbatim}

\begin{verbatim}
www.iaa.es/~ortiz/onlinefission4.avi
\end{verbatim}

\section*{Acknowledgments}

 This research was partially supported by Spanish
grants AYA2008-06202-C03-01, AYA-06202-C03-02, AYA2008-06202-C03-03,
P07-FQM-02998 and European FEDER funds.
RD acknowledges financial support from the MICINN (contract Ram\'on y
Cajal). DCR acknowledges support from the National Aeronautics and Space
Administration under grant No. NNX08AM39G issued through the office of Space
Science. PGB took part in this research during her stay at the Southwest
Research Institute (Boulder, CO, USA) with support from the Ministerio de
Ciencia e Innovaci\'on (Spain). We thank Seth Jacobson and Daniel Scheeres
for sharing with us their manuscript in preparation.

\end{document}